\documentclass{aa}  
\usepackage{graphicx}
\usepackage{txfonts}
\begin{document} 
   \title{Asteroseismogyrometry of low-mass red giants}

   \subtitle{I. The SOLA inversion method}

   \author{F. P. Pijpers
 \inst{1}
          \and
          M. P. Di Mauro\inst{2}
          \and R. Ventura\inst{3}
         } 

   \institute{Korteweg-de Vries Institute for Mathematics, University of 
              Amsterdam, Science Park 105-107, 1098 XG Amsterdam, Netherlands
              \email{f.p.pijpers@uva.nl}
         \and
             INAF-IAPS Istituto di Astrofisica e Planetologia Spaziali, Via del Fosso del Cavaliere 100, 00133 Roma, Italy\\
          \email{maria.dimauro@inaf.it}
             \and 
             INAF-Osservatorio Astrofisico di Catania,
            Via Santa Sofia 78, 95123 Catania, Italy\\ \email{rita.ventura@inaf.it}
             }

   \date{Received; accepted }

 
  \abstract
   {During the past 10 years the unprecedented quality and frequency resolution of asteroseismic data provided by space photometry have 
revolutionised the study of red-giant stars providing us with the possibility to probe the interior of  thousands of these targets.}
   {Our aim is to present an asteroseismic tool which allows to determine the total angular momentum of stars, without a priori inference  of their internal rotational profile. }
   {We adopt and adapt to red giants the asteroseismic inversion technique developed for the case of the Sun. The method has been tested assuming different artificial sets of data, including also modes with harmonic degree $l\geq 2$.}
   {We estimate   with an accuracy of $14.5\%$ the total angular momentum of the red-giant star KIC~4448777 observed by \textit{Kepler} during the first four consecutive years of operation.}
   {Our results indicate that the measurement of the total angular momentum of red-giant stars can be determined with a fairly high precision by means of asteroseismology by using a small set of rotational splittings of only dipolar modes and that our method, based on observations of stellar pulsations, provides a powerful mean for testing and modelling transport of angular momentum in stars. }

   \keywords{asteroseismology -- red-giant stars --pulsations --rotation -- angular momentum}

   \maketitle
%

\section{Introduction}
The total angular momentum of a star is a fundamental quantity directly related to its internal structure and rotation rate  \citep[e.g.,][]{Maeder2009}. Stars lose a significant amount of angular momentum, between their initial formation and the final stages,  and the processes transporting it  in their interior
play a key role in stellar  evolution, producing mixing of chemical species and modifying the structure and the chemical gradient between the surface and the core of the star during its life.
However, questions related to the transport of angular momentum inside the stars are not yet fully understood: several mechanisms seem to be active, but they are only approximately or just not modelled at all 
\citep[see e.g.,][]{GoughMcIntyre1998, Spruit1999, Spruit2002, CharbonnelTalon2005, Spadaetal2010, MaederMeynet2012, Mathis2013, Zahn2013, Eggenbergeretal2017, Aerts2019}. 

The study of the stellar total angular momentum is also crucial for a better understanding of the dynamical histories of multi-planet systems and modelling of angular momentum distribution and exchange within host-star and planets. This question, quite controversial and  widely discussed in literature \citep[e.g.,][]{Alvesetal2010, Paz2015, Irwin2015},
is dealing at the moment with several hypotheses including a possible
correlation between exoplanetary masses and orbital distances \citep[e.g.,][]{shashanka2019} and a connection between the multiplicity of the systems and their dynamical history \citep{zinzi2017}.

In short, the knowledge of the global angular momentum of a star is critical to understand, test and improve theories able to describe the angular momentum prescriptions in various  astrophysical contexts. 

The total angular momentum of a star is, however, a quantity inaccessible to direct observations; what one can measure is only the projected surface rotational velocity obtained by stellar spectroscopy, thus it has been evaluated so far just by adopting methods referring  to the stellar fundamental parameters  and power law relationships, regardless of the real distribution of mass and angular momentum inside  the stellar interior \citep[see, e.g.,][] {Irwin2015}. 

Over the last decade the extremely successful observations of stellar pulsations of unprecedented  high quality  by the space missions CoRoT \citep{Baglinetal2006} and Kepler  \citep{Boruckietal2010}, and even more recently, by the current NASA explorer TESS satellite \citep{Rickeretal2014},  opened up enormous opportunities and perspectives  for improving our  knowledge of  otherwise hidden stellar interiors and shedding new light on stellar evolution. 

In particular, the observation of stellar pulsations has enabled the measurement of the internal rotation in a large sample of stars with different masses and evolutionary stages, by using the  splittings of the oscillation frequencies caused by rotation in  their power spectra.  
This opportunity is allowing us to put the necessary constraints on stellar angular momentum theories: it has been shown that a strong decrease of core’s rotation occurs during stellar evolution irrespective of the star's mass or binarity  \citep{Aerts2019}, with values at least two orders of magnitude lower than currently theoretically predicted. This is a strong element  indicating that one or several mechanisms capable of extracting angular momentum from the core must be at work during stellar evolution
\citep[e.g.,][]{marques2013, goupil2013,cantiello2014,ouazzani2019}. 
Moreover, the efficiency of these angular momentum transport processes, still unknown, appears mass dependent \citep{Eggenbergeretal2017}, although it is not clear if the angular momentum transport increases gradually in efficiency with stellar mass.
 
Among the asteroseismic targets, the red-giant branch stars (RGB) are of particular interest in this context, showing dense oscillation spectra characterised
by the presence of modes with both gravity and acoustic character, known as "mixed modes", able to probe the physical  conditions from the inner core to the envelope 
\citep[e.g.,][]{Becketal2012, Deheuvelsetal2012, Mosseretal2012, Deheuvelsetal2014, DiMauro2016}.
The results, so far, have shown the presence in low mass red-giant stars of rapidly rotating helium cores up to 10 times faster than the envelope  and  shear layers located between the core and the hydrogen burning shell \citep{DiMauro2018}. 

Here, we present a method based on the asteroseismic inversion technique developed by \cite{Pijpers1998}, which allows us to compute the total angular momentum of a star by avoiding the need to infer first the stellar rotation rate by the inversion of the observed  rotational splittings. This has the great advantage of drastically reducing  the number of necessary computational steps and, as well, the systematic errors introduced at each step, ultimately improving the precision and the accuracy of the results. 

In the case of the Sun, the total angular momentum deduced by applying the so called SOLA inversion technique  to  $414$ acoustic modes  by MDI/SOHO, with harmonic degree $1\leq l\leq 250$, is $\overline{J_\odot} = (1.92 \pm 0.019) \cdot\, 10^{48} \mathrm{g\, cm^2\, s^{-1}}$ as obtained by \citet{Pijpers1998}.  More recently  \cite{Pijpers2003} demonstrated that it is still possible to obtain an accurate estimate (with an uncertainty of about 3\%) of the total angular momentum of the Sun even by inverting a small set of only 14 modes  with $l=1$.  This is not so surprising if we look at the nature of solar acoustic modes, those in which the Sun is observed to oscillate predominantly, and showing amplitudes high enough to be detected at the surface. They, in contrast to gravity modes, which in the Sun are trapped in the radiative interior and evanescent in the convection zone,  are not very sensitive to the physical structure of the solar core because their energy density is inversely proportional to the sound speed, which is quite large in the solar radiative region. Only low order, low harmonic degree ($l\leq4$) p-modes show energy distributed throughout most of the Sun, providing us a chance of sensing the physical conditions of the solar deepest interior. The result obtained by \citet{Pijpers2003}  could pave the way for the SOLA technique to be applied to determine the total angular momentum also in stars other than the Sun, where a small number of modes are detectable in their power spectra. 

 The aim of this work is to investigate how reliable it could be to estimate the total angular momentum of red-giant stars with a fairly good accuracy even from a small set of data (presumably limited to dipole modes) by applying an adaptation of the SOLA method and taking advantage of the high-precision photometry provided by the most recent space missions.
 Of course high-precision does not necessarily translate into accuracy, thus
   we will distinguish the statistical error, rising from the uncertainties in the data, from the systematic error arising from the constraints in stellar mass and radius.

 In this context the red-giant star KIC~4448777, observed by the Kepler satellite, offers a promising test case. The star, located at the beginning of the ascending red-giant branch, has been observed for more than four years, during the Kepler satellite’s first nominal mission and has been deeply investigated by \cite{DiMauro2016, DiMauro2018}. The authors were able to identify  20 rotational splittings of mixed modes  over a total of 77 individual observed frequencies and to infer the details of its internal rotational profile. Starting from the resolved rotation rate and the best-fit evolutionary model of the star computed by \cite{DiMauro2016, DiMauro2018}, we first derived the value of the total angular momentum  by just solving the proper integral equation.  Then, we re-computed the global angular momentum of the star by applying the asteroseismic inversion technique proposed by \cite{Pijpers1998, Pijpers2003}  to the 20 observed splittings  and compared the results, thus obtaining at last a test of this method in the case of red-giant stars. 

The paper is structured as follows: Sect. 2 provides details on the determination of the total angular momentum in stars and presents the methods for calculating the total angular momentum through the use of asteroseismic data. Sect. 3 tests the inversion method and shows the results obtained for KIC~4448777. Sect. 4  discusses the results and draws some conclusions. 

\section{The total angular momentum}
\subsection{Basic equations}
The total angular momentum of a star is a global fundamental parameter, related to the internal rotation rate and density distribution in the stellar interior  through an  integral equation:
\begin{equation} 
 J_{\mathrm{tot}}=\int_0^1 \int_{-1}^1 I(x,u)\frac{\Omega(x,u)}{2\pi} \,\mathrm{d}u\,\mathrm{d}x,
\label{Eq.1}
\end{equation}
where $x = r/R $ is the fractional radius,  $R$ the radius of the star,  $u=\cos \theta$  with  $\theta$  the co-latitude, $\Omega(x,u)$ is the internal rotation rate and  $I(x,u)$ is the scalar moment of inertia,  critically dependent on how the stellar mass is distributed inside the star and given by:
\begin{equation}
\label{eq:angmomker}
 I(x,u)=2\pi R^5\rho(x) x^4(1-u^2)=I(x) \frac{3}{4}(1-u^2) 
\end{equation}
where $\rho(x)$ is the density as function of the fractional radius inside the star.

As a  first-order approximation the angular velocity in the inner regions of a star can be
considered  almost independent of latitude. 
As a consequence, integrating Eq.~(\ref{Eq.1}) over  $\cos\theta$, after trivial calculations, produces:
 \begin{equation}
 J_{\mathrm{tot}}=\int_{0}^1J(x) \mathrm{d}x=
\int_{0}^1I(x) \frac{\Omega(x,u)}{2\pi}\mathrm{d}x,
\label{Eq.3}
\end{equation}
where the total moment of inertia is
\begin{equation}
\label{Eq.4}
I_{\mathrm{tot}}=\int_{0}^1 I(x) \mathrm{d}x=\frac{8}{3}\pi R^5\int_{0}^1x^{4}\rho(x) \mathrm{d}x.
\end{equation}
The main  difficulty to compute this integral in real stars comes from the impossibility to know a priori, from classical observations, both the mass distribution inside their interior, namely the dependence of $\rho$ on the fractional radius, and the full radial dependence of the rotation rate profile.

\subsection{Computation based on known internal rotational profile}
A rough estimate of  $ J_{\mathrm{tot}}$ in stars has been obtained, so far, by calculating  the moment of inertia over a stellar structure model and adopting the observed  projected surface rotation velocity as an approximation of the average internal angular velocity, actually neglecting any dependence on the fractional radius in the interior of the star.

Two  fundamental  questions arise: i) how representative of the true internal rotation is the surface value adopted \citep[e.g.][] {Benomaretal2015, Saioetal2015} and ii) what should be the consequence  of the adopted approximation on the  accuracy of the results.

In the case of the Sun, in which the deviations from an uniform rotation in the interior are relatively small, the total angular momentum obtained from computations based on the observed surface rotation is  $ J_{\mathrm{surf}} = 1.63\cdot  10^{48} \mathrm{g\, cm^2\, s^{-1}}$ \citep{Cox2000},
a value which is about 14\% smaller than that deduced by integrating  the internal rotational profile obtained by helioseismic inversion of the MDI/SOHO data $J_{\odot}= (1.96\pm 0.05) \cdot 10^{48} \mathrm{g\, cm^2\, s^{-1}}$  \citep{DiMauro1998}.
 
Up to now, we do not have other direct experience of these kind of studies applied to stars more evolved or more massive than the Sun, but we expect even larger discrepancies for the case of distant stars.
Nevertheless, we suppose that for any significant study on the dependence of the angular momentum on stellar mass and age a high precision should be required.

\subsection{The SOLA inversion method}

The total angular momentum of a star, as demonstrated by \cite{Pijpers2003}, can be determined by the asteroseismic inversion of the following integral equation: 
\begin{equation}
\delta \nu_{i}= \int_{0}^{1}\mathrm{d}x\int_{-1}^{1}J(x,u)\, \mathrm{d}u+ \epsilon_{i}\;.
\label{eq:rot}
\end{equation}

This equation which relates a set of $N$ observed
 rotational splittings $\delta \nu_{i}$ with uncertainties $\epsilon_{i}$,
for the modes $i\equiv(n,l,m)$ of radial order $n$, harmonic degree $l$ and azimuthal order $m$
  to the internal angular momentum profile $J(x,u)$,
  is
derived from the application of a standard perturbation theory to an
equilibrium stellar structure model, under the hypothesis of slow rotation \citep{Pijpers1994}.

 In fact,
the rotation breaks the spherical symmetry of the stellar
structure and splits the central frequency of each oscillation mode of
harmonic degree $l$ in a multiplet with $2l+1$ components separated by a frequency
splitting defined by:
\begin{equation}
  \delta \nu_{i}=  \frac{\nu_{nlm}-\nu_{nl-m}}{2m} \, ,
 \end{equation}
where for notation convenience the single index $i$ is used to
enumerate the available multiplets $nl$.

As demonstrated by \cite{Pijpers2003}, the 2-dimensional expression (Eq.~\ref{eq:rot}) can  be
simplified 
in the following equation:
\begin{equation}
\delta \nu_{i} = \int_{0}^{1}\mathrm{d}x\,{\cal K}_{i}(x) \int_{-1}^{1}\frac{3}{4}(1-u^2)\frac{ \Omega(x,u)}{2\pi}\, \mathrm{d}u +\epsilon_{i}\; .
\label{eq:mom}
\end{equation}
The individual kernels ${\cal K}_{i}(x)$  are function of the radius only (cf. \citealp{Pijpers1997, Pijpers2006}) and are calculated on the unperturbed eigenfunctions of the modes
and other physical quantities of the stellar model which best reproduces all the observational constraints of the star:
\begin{equation}
{\cal K}_{i}(x)=\rho x^2\left[\xi_{nl}(x)^2 - 2\frac{1}{L}\xi_{nl}(x) \eta_{nl}(x) +\left(1-\frac{1}{L^2}\right)\eta_{nl}(x)^2 \right] / {\cal{I}}_{nl}\, ,  \nonumber
\end{equation}
where
$\xi_{nl}(x)$ and $\eta_{nl}(x)$ are the radial and horizontal components of the displacement eigenfunctions of the modes, $L^2=l(l+1)$, and ${\cal{I}}_{nl}=\int_0^1  \rho \left[\xi_{nl}(x)^2+\eta_{nl}(x)^2\right]\mathrm{d}x$
is a normalisation factor which ensures that all kernels integrate to $1$ over their domain. 
 This means that the determination of the angular momentum takes the full 2-dimensional character of rotation into account, but nevertheless requires evaluating only 1-D integrals.
There is also no need for correcting for the angle of inclination of the rotation axis with respect to the line of sight.

The integral Eq.~(\ref{eq:mom}) can be solved
 by applying the Subtractive Optimally Localized Averages (SOLA) method \citep{Pijpers1992,Pijpers1994}.
The SOLA method is designed to provide a weighted average of
 a given function
by means of a  linear combination of all the data, so that:
\begin{equation}
\overline{J_\star}=\sum_{i=1}^{N}c_{i}\delta\nu_{i}=
\int_{0}^{1}\mathrm{d}x K(x)\int_{-1}^{1}\frac{3}{4}(1-u^2)  
\frac{\Omega(x,u)}{2\pi}\mathrm{d}u + \sum_{i=1}^{N}c_{i}\epsilon_i\; ,
\label{backus}
\end{equation}
where $c_{i}$ are the inversion coefficients,
and 
\begin{equation}
K(x)=\sum_{i=1}^{N} c_{i}{\cal K}_{i}(x)
\label{Fker}
\end{equation}
 is the averaging kernel. 
 
For solving 
the inversion, one should construct a linear combination of kernels, that resembles "as closely as possible"
an appropriate chosen target function ${\cal T}(x)$.
This is possible, because
compared to the older inversion method by \cite{Backus70}, also often referred to as OLA, the SOLA method has a main advantage:  the possibility to choose any target function that achieves a certain purpose. 
In the case one wishes to determine a resolved local measurement of the rotation rate, the target function is usually chosen to
be a Gaussian, centred on that location. 
However, choosing a Gaussian function, while a common choice, is not essential to the method. 
As demonstrated by \cite{Pijpers2003} for measuring the total angular momentum of a star the
appropriate chosen target function is well represented by
the moment of inertia $I(x)$ (see Eqs.~\ref{Eq.1} and \ref{eq:angmomker}).

As in every inverse method, regularisation is necessary to find an acceptable matching of the averaging kernel
to its target function and also to ensure an acceptable small error on the result arising
from the propagation of the data errors. In the SOLA method this regularisation takes the form of balancing on one hand the minimisation of the squared difference of the linear combination of kernels and the target kernel, and on the other hand minimising the propagated measurement errors. 
For a given target function ${\cal T}(x)$, the expression for the functional ${\cal F}$ that should be minimised is:
\begin{equation}
\label{eq:functional}
{\cal F} \equiv \int\limits_0^1\left[\sum\limits_{i} c_{i}{\cal K}_{i}(x) -{\cal T}(x)\right]^2 \mathrm{d}x + \mu\sum\limits_{ij} c_{i} E_{ij} c_{j}.
\end{equation}
This should be minimised for the linear coefficients $c_{i}$ by adjusting $\mu$, which is an error weighting parameter, while the matrix $E_{ij} $ is the errors variance-covariance matrix for the splittings data. 
While there are cross-validation strategies intended to
find an optimal value for $\mu$, in practice some experimentation can be useful or even necessary. 
Note that when minimising the functional ${\cal F}$ of Eq.~(\ref{eq:functional}) for the coefficients $ c_i$, it is usual to enforce that the sum of the coefficients be equal to $1$ using a Lagrange multiplier $\bf v$. 
 Hence, the inversion problem is equivalent to solve
the following set of linear equations \citep[see][]{Pijpers1994}:
\begin{equation}
\mathrm{W}{\bf c}={ \bf v}\; , 
\label{vect}
\end{equation}
where $\mathrm{W}$ is the symmetric $(N+1)\times (N+1)$ mode cross-correlation matrix
whose elements are
$W_{ij}= \int\limits_0^1{\cal K}_{i}(x){\cal K}_{j}(x)\mathrm{d}x +\mu E_{ij}$ and
 ${\bf v}$ is the vector which includes the  cross-correlation eigenvalues of the kernels with the target function ${\cal T}(x)$:
 \begin{equation}
 v_i= \int\limits_0^1{\cal K}_{i}(x){\cal T}(x)\mathrm{d}x \,.
 \end{equation}
 The solutions of Eq.~(\ref{vect}) are the inversion coefficients $c_i$ collected in the vector $\bf c$,
 obtained by inverting the matrix $\mathrm{W}$ only one time.

\section{Asteroseismogyrometry of the red-giant KIC 4448777}
\subsection{Computation by integrating the angular velocity profile}
\label{sec3.1}
KIC~4448777, located at the beginning of
the ascending red-giant branch, is a solar-like star, where oscillations stochastically excited by turbulent convection, as in the Sun, have allowed \cite{DiMauro2016, DiMauro2018} to fully characterise its structure.
Having exhausted its central hydrogen supply, it has a degenerate helium core surrounded by a still burning hydrogen shell. Table \ref{tab_fitted} reports the main characteristics of the star.  In particular,
the effective temperature, the gravity and iron abundances $[\mathrm{Fe/H}]$
 were obtained from spectroscopic campaigns.
 The large separation $\Delta \nu$, reported in Table \ref{tab_fitted}, was calculated by linear fit over the asymptotic relation for the observed radial mode frequencies \cite[see][for more details]{DiMauro2018}.
  The asteroseismic mass $M=(1.12\pm0.09) M_\sun$ and  radius 
$R=(4.13\pm 0.11) R_\sun$ were obtained directly by the asteroseismic scaling-laws based only on the spectroscopic atmospheric parameters and the asteroseismic measurements of $\Delta \nu$ and $\nu_{max}$, respectively the observed large-separation  and  frequency of the maximum amplitude power.
 
 Table \ref{tab_fitted} shows also all the theoretical parameters of the best model, including
 the luminosity $L$,  the age,  the location of the base of the convective region $r_{cz}/R$, the extent of the He core $r_{He}/R$  of the stellar reference model 
selected in \cite{DiMauro2016, DiMauro2018} as the
result of the best-fitting procedure between KIC~4448777 and hundreds of evolutionary models built in order to match simultaneously all the observational spectroscopic and seismic constraints, including all the 77 individual observed frequencies with harmonic degree $l=0,1,2,3$. This is part of a standard and well developed procedure for all the asteroseismic analysis, which ensures that the chosen reference model provides the closest possible description of the real star.
In particular, this selection is necessary for application of any inversion method which requires linearization
around a reference model close enough to the actual star in order to provide reliable results. The procedure of selection was largely discussed in \cite{DiMauro2016,DiMauro2018}.

\begin{table} [ht!]
\caption{Main observed and theoretical parameters for KIC~4448777.} 
\centering
\begin{tabular}{lcc}
\hline
\hline
 & KIC~4448777 & Best-fit model\\
 \hline 
 $T_{\mathrm{eff}}$ (K) & $4750\pm250$\tablefootmark{a} &    4735  \\
 $\log g$ (dex) & $3.5\pm0.5$\tablefootmark{a} &   3.27 \\
 $[\mathrm{Fe/H}]$ & $0.23\pm 0.12$\tablefootmark{a} & $0.13$\\
 $\Delta \nu\, (\mu \mathrm{Hz})$ & $16.973\pm0.008$\tablefootmark{b}  & $16.933$\\
 $\nu_{\mathrm{max}}\,(\mu \mathrm{Hz})  $& $219.75\pm1.23$\tablefootmark{b}& -\\
 $M/{\mathrm M}_{\odot}$ & $1.12\pm0.09$\tablefootmark{b}  & 1.13  \\

 $R/{\mathrm R}_{\odot}$ & $4.13\pm 0.11^b$ &  4.08 \\
 Age (Gyr) & -& 7.24  \\
 $L/{\mathrm L}_{\odot}$ & -  &   7.22  \\

$r_{cz}/R$& - &  0.1448 \\ 
$r_{He}/R$ & -  & 0.0074 \\

\hline 
&&\\
\end{tabular}
\label{tab_fitted}
\tablefoot{  \\
\tablefoottext{a}{Determined by spectroscopic observations \citep[see][]{DiMauro2016}}
\tablefoottext{b}{Determined by asteroseismic observations\citep[see][]{DiMauro2016,DiMauro2018}}
}
\end{table}

The star has been observed by
the {\it Kepler} satellite for more than four years, during
its first nominal mission and its internal rotation has been successfully determined by means of asteroseismic inversion \citep{DiMauro2018}.
The detection of 20 rotational splittings of mixed modes allowed the authors to reconstruct in details the angular velocity profile into the interior of the star, showing that the helium core rotates almost rigidly about 6 times faster than the convective envelope, while part of the hydrogen
shell seems to rotate at a constant velocity, about 1.15 times lower than the He core.

\begin{figure}
\centering
\includegraphics[width=9cm]{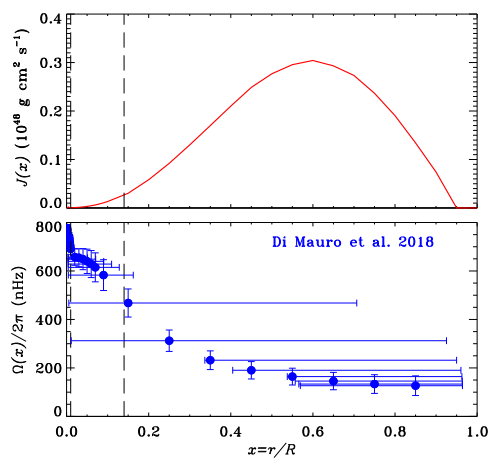}
\caption{Upper panel: radial profile of the angular momentum  inside the best-fit model of KIC~4448777  according to the definition of Eq. (\ref{Eq.3}). Lower panel: radial  profile of the angular velocity  as obtained by inversion of rotational splittings in \cite{DiMauro2018}. Dashed lines show the base of the convective zone $r_{cz}=0.1448\,R$ and the external edge of the He core $r_{He}=0.0074\,R$.}
    \label{angmom}
\end{figure}

The total angular momentum of KIC~4448777 obtained by integrating
Eq.~(\ref{Eq.3}) and using $\Omega(x)$ determined by the inversion of the rotational splittings in \cite{DiMauro2018}
is $J_{\mathrm{tot}}=3.00^{+5.8}_{-1.5}\cdot 10^{48}\,\mathrm{ g\,cm^{2}\,s^{-1}}$  and the moment of inertia integrating Eq.~(\ref{Eq.4}) is $I_{\mathrm{tot}}=1.85 \cdot 10^{55}\,  \mathrm{g\,cm^{2}}$. The large error in $J_{\mathrm{tot}}$ reflects the low spatial resolution with which the angular velocity can be determined  in the regions relevant for the determination of the angular momentum, as shown in  Fig.~\ref{angmom} which reports 
the radial dependence of the angular momentum  computed  for  the best-fit model of KIC~4448777 and the radial profile of the rotation rate as inferred in \cite{DiMauro2018}.

The contribution to the total angular momentum inside the star comes mainly from the layers of the convective envelope between $r=0.2\, R$  and $r=0.9\,R$ (see Fig. \ref{angmom}),  where the uncertainties in the angular velocity are higher, while the remaining part of the star is almost unimportant, despite the high velocity rate in the inner regions. 

\begin{figure}
    \centering
    
     \includegraphics[width=9cm]
    {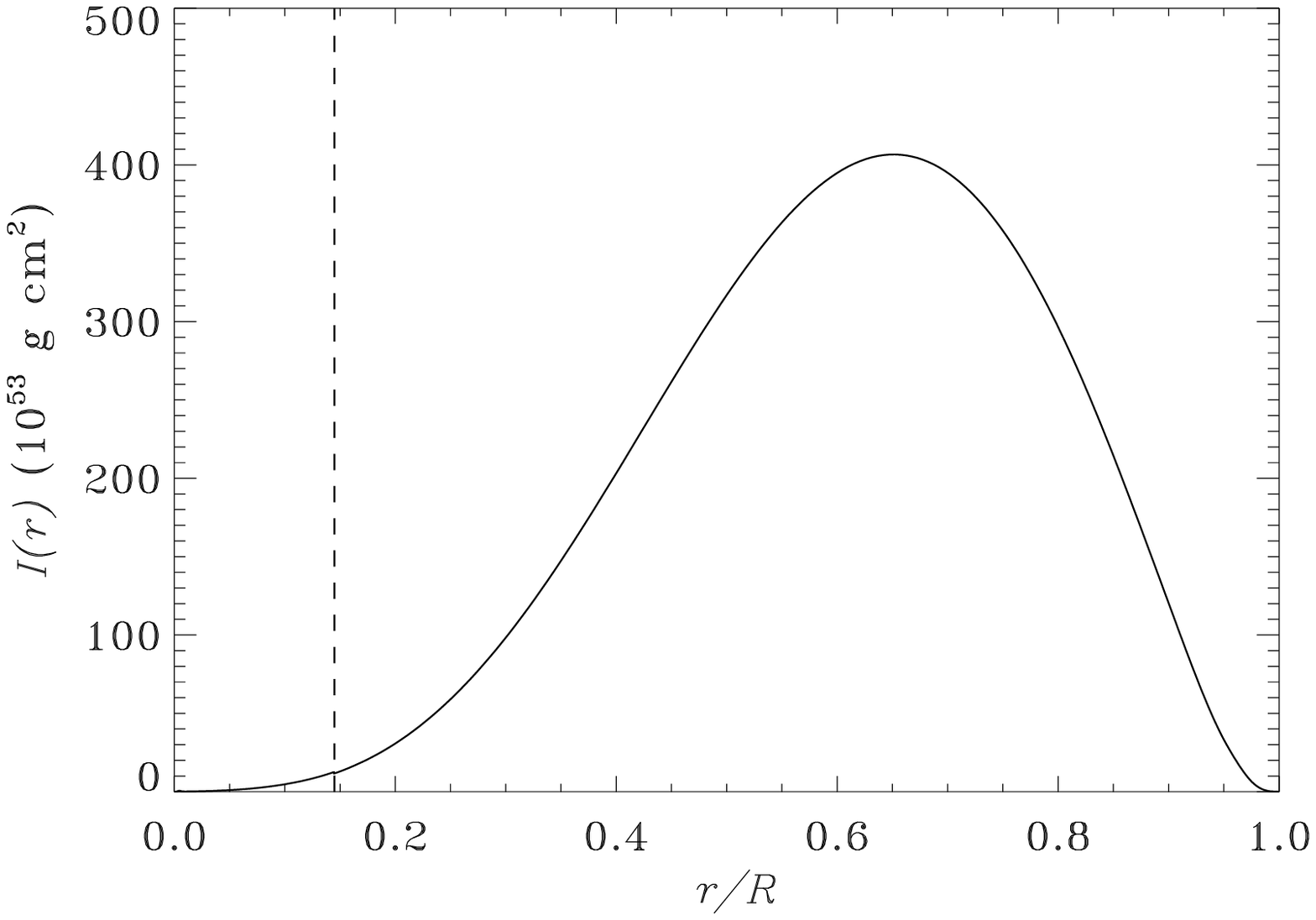}
    \includegraphics[width=9cm]
    {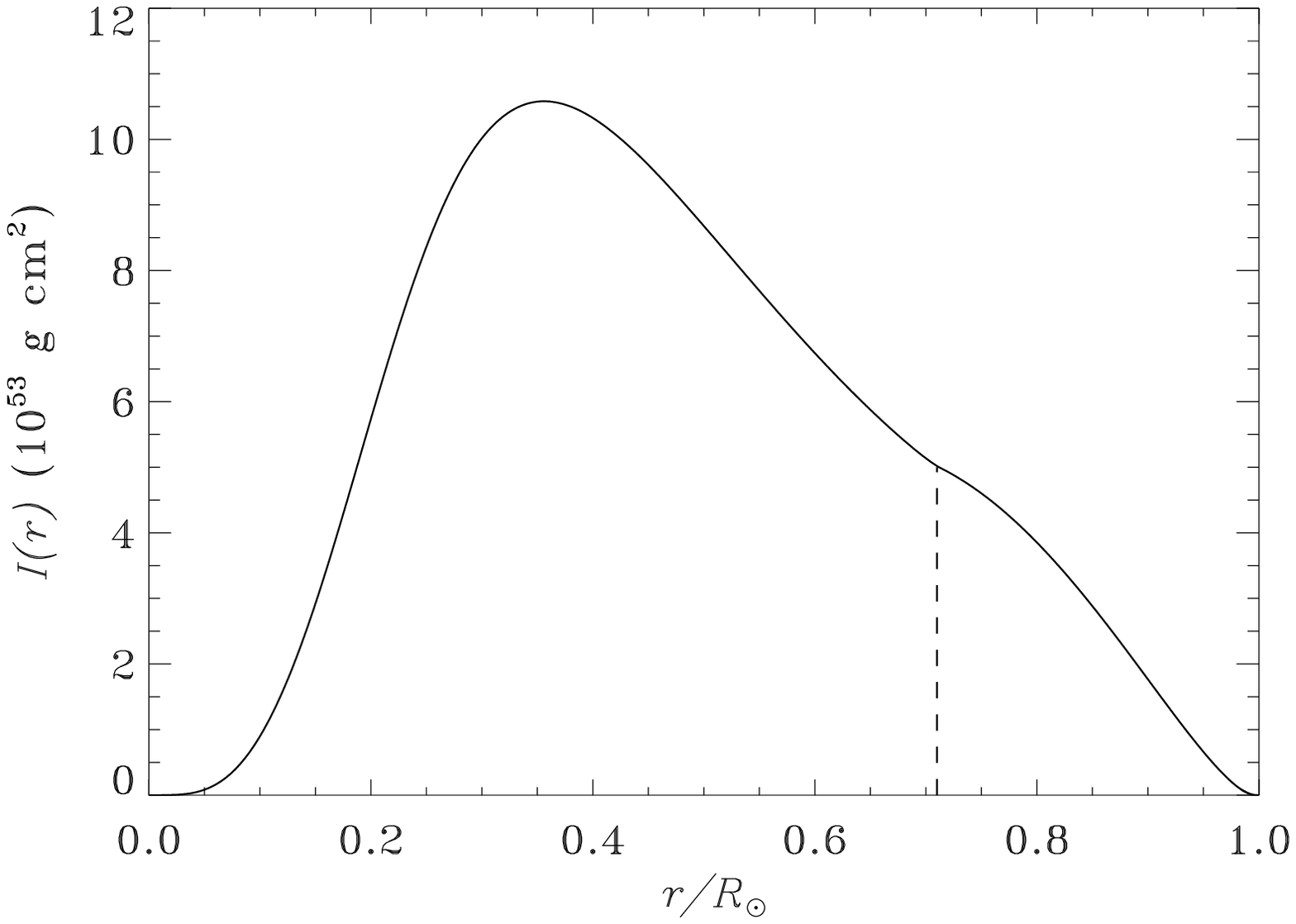}
    \caption{Radial profile of the moment of inertia inside the best-fit model of KIC~4448777 (top panel) and - for comparison - inside model S of \citet{jcd1996} for the Sun (low panel). The location of the base of the convective zone is shown by dashed lines.}
    \label{inertia}
\end{figure}

This result is  mostly determined by the radial distribution of the moment of inertia  inside the star, as can be seen in  Fig. \ref{inertia}, where  the moment of inertia $I(x)$  calculated for our best-fit model of KIC~4448777 and, for comparison, for model S of the Sun \citep{jcd1996} are shown. 
By simply considering that the total moment of inertia can be split in two integrals calculated over the two adjacent regions, the convective envelope and the radiative layer,
\begin{equation}
I_{\mathrm{tot}}=\int_{0}^{1} I(x)\mathrm{d}x=
\int_{0}^{r_{cz}} I(x) \mathrm{d}x+
\int_{r_{cz}}^{1} I(x) \mathrm{d}x=I_{rad}+I_{cz},
\end{equation}
we can try to roughly understand which region is mostly contributing to the moment of inertia as a solar-type star evolves.
The respective  values of $I_{cz}$ and
$I_{rad}$ given in 
Table \ref{tab.inertia} show that, at odds with the Sun, only the convective envelope is contributing significantly to the total moment of inertia of KIC 4448777:
the core and the H-burning shell have a very high density, but strongly confined in a too small fraction of the stellar radius, while the upper layers in the red giants are far too tenuous to be able to supply substantial contributions.
\begin{table} [h!]
\caption{Moment of inertia inside the convective envelope  $I_{cz}$ and in the radiative region $I_{rad}$ calculated for our best-fit model of KIC~4448777 and for a solar model.} 
\centering
\begin{tabular}{lccc}
\hline
\hline
 & $I_{rad} (\mathrm{g\,cm^{2}}) $& $I_{cz} (\mathrm{g\,cm^{2}})$&$I_{\mathrm{tot}}
( \mathrm{g\,cm^{2}})$\\
 \hline 
 KIC 4448777&   $6.95 \cdot 10^{52}$&$1.84  \cdot 10^{55}$ &$1.85 \cdot 10^{55}$\\
 Sun &  $4.56  \cdot 10^{53}$& $7.63\cdot 10^{52}$ & $5.32\cdot 10^{53}$\\
\hline 
&&\\
\end{tabular}\\
\label{tab.inertia}
\end{table}

We will show in the next sections that the direct asteroseismic method suggested by \cite{Pijpers2003} provides a more powerful tool to determine accurate estimates of the total angular momentum also in stars other than the Sun, without the need to solve a priori the inner rotational profile. 

\subsection{Testing the SOLA inversion procedure}

In order to test the potential of the SOLA procedure for the case of red giants, we carried out a 'hare-and-hounds' exercise, in which the hare is represented by the total angular momentum calculated for the best-fit model of KIC~4448777 summarised in Table \ref{tab_fitted} and assuming
 a simple fictitious rotational profile given by:
\[
\left\{\begin{array}{ll}\Omega(r)=750\, \mathrm{nHz} &  r\leq 0.05R\\ \Omega(r)=120\, \mathrm{nHz}&  r>0.05R\\
\end{array} \right.
\]
This rotational profile, by solving 
Eq.~(\ref{Eq.3}), produces an angular momentum $J_{(1)}= 1.49 \cdot 10^{48} \mathrm{g\, cm^2\, s^{-1}}$.

Using a forward seismological approach, as described in \citet{DiMauro2016},  we computed a set of 20 artificial rotational splittings for $l=1$ from the above step-like rotational law.
To get a realistic
estimate of the capabilities of the inversion,
this synthetic set of data comprises
 the same modes observed for KIC 4448777 and for each rotational splitting
  we adopted the same
 error of 15 nHz, an average of the errors of the  observed data (between $0.5\, \mathrm{nHz}\leq \epsilon\leq 30\,\mathrm{nHz}$).

The hounds are the values of angular momentum $\overline{J_{(2)}}$ obtained by the attempts at inversion
of the set of  artificial splittings by varying the trade-off parameter $\mu$ in order to determine the closest value to
$J_{(1)}$.

As explained in Sect. 2, because of the ill-conditioned nature of the inversion problem, we should apply a regularisation procedure by varying the trade-off parameter $\mu$ in order to choose a good compromise
between uncertainty of the solution and mismatch between the averaging kernel ${ K}(x) $ and the target kernel ${\cal T}(x)$ given by:
\begin{equation}
    \chi^2=\int_0^1 \left[\sum\limits_{i} c_{i}{\cal K}_{i}(x) -{\cal T}(x)\right]^2 \mathrm{d}x\,.
\end{equation}
The choice of the value of $\mu$, which depends strongly on the magnitude of the errors  in the data and on the oscillation modes adopted, sets the trade-off between
accuracy and precision of the solution.
By lowering the trade-off parameter it is possible to obtain averaging kernels closer to the target kernels,  which in other words means a lower $\chi^2$ value, but this decreases the precision with which the solution is determined, since the weight of the errors increases.  Thus,
we should choose, among all the possible solutions, the one obtained for a value of $\mu$ which provides an optimal compromise between low $\chi^2$  and  a small error in the result. 
 
\begin{figure*}
    \centering
    \includegraphics[width=9cm]{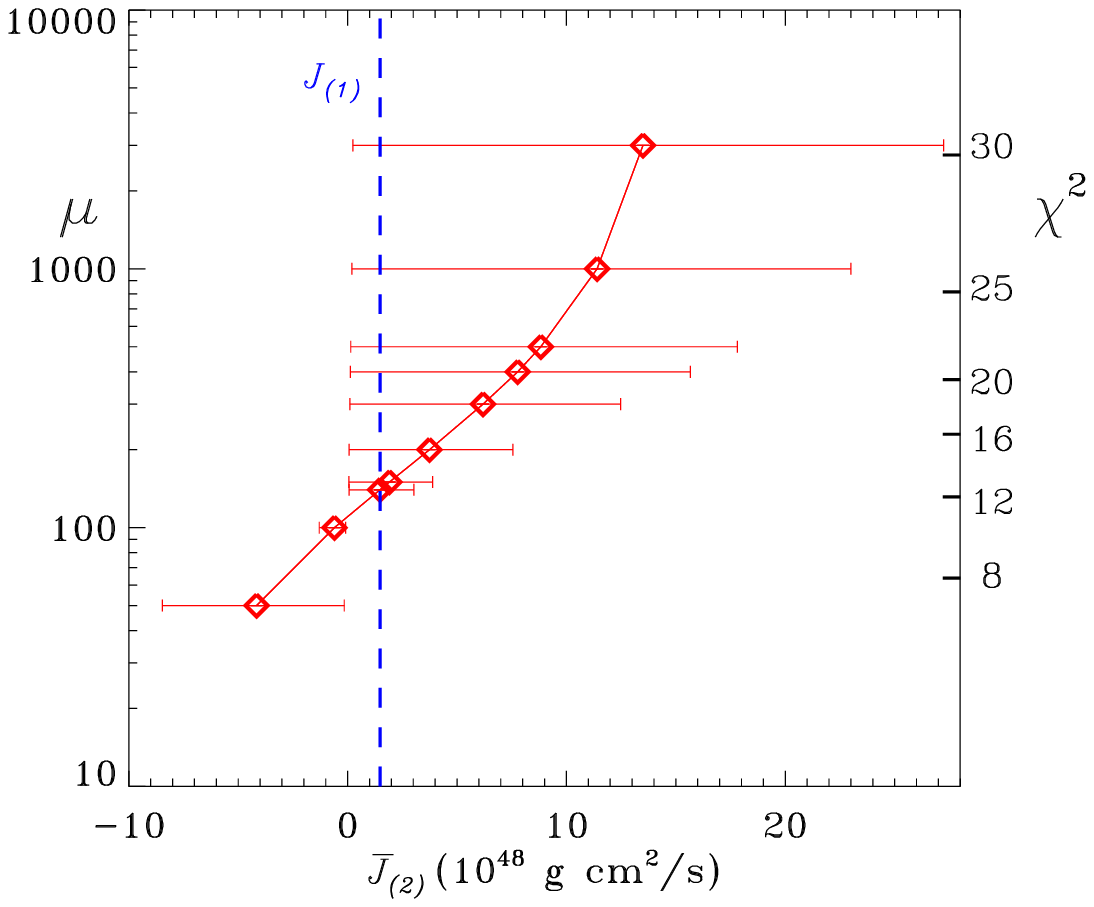}
     \includegraphics[width=9cm]{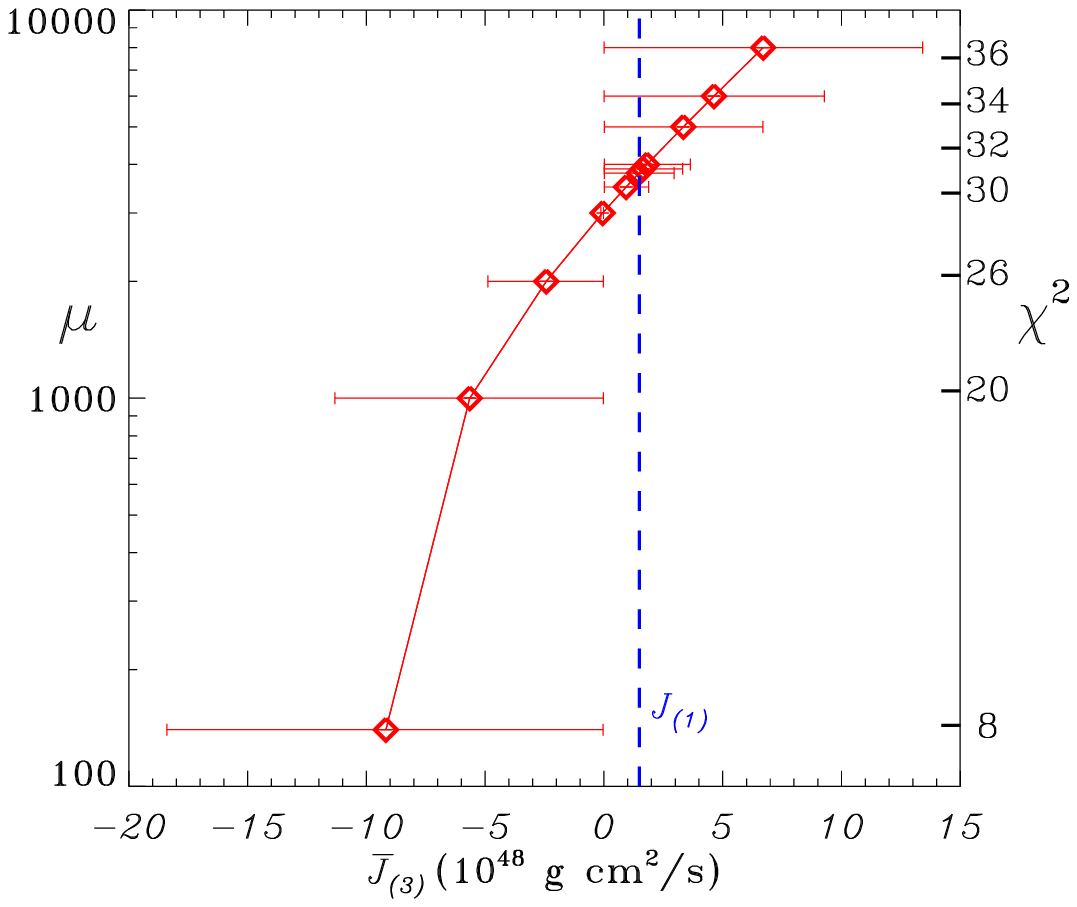}
  \caption{Hare and hound attempts for a set of synthetic data including  the same modes as those observed for KIC~4448777. Inversion solutions (red diamonds) are plotted as function of the trade-off parameter $\mu$ and the relative values of $\chi^2$.
  The left panel shows the results for a set of splittings with errors  $\epsilon_i=15\, \mathrm{nHz}$. The right panel shows the results for the same synthetic set assuming real observed errors as in \citet{DiMauro2018}.
  The theoretical value ${J}_{(1)}=1.49 \cdot 10^{48} \mathrm{g\, cm^2\, s^{-1}}$, which should be matched by the inversion procedure, is shown by the blue dotted line.}
    \label{chi-mu}
\end{figure*}

By varying the trade-off parameter between $100\leq \mu\leq 3000$, we obtained for $\mu=140$ the solution $\overline{J_{(2)}}=(1.48\pm 0.07) \cdot 10^{48} \mathrm{g\, cm^2\, s^{-1}}$ characterised by $\chi^2=12$, which represents the best agreement with $J_{(1)}$.
This is well shown in the trade-off diagram of Fig. \ref{chi-mu} (left panel), where the variation of the total angular momentum $\overline{J_{(2)}}$ is plotted against various values of $\mu$, while the values of $\chi^2$ are shown on the ordinate axis on the right.

\begin{figure}
    \centering
    \includegraphics[width=9cm]{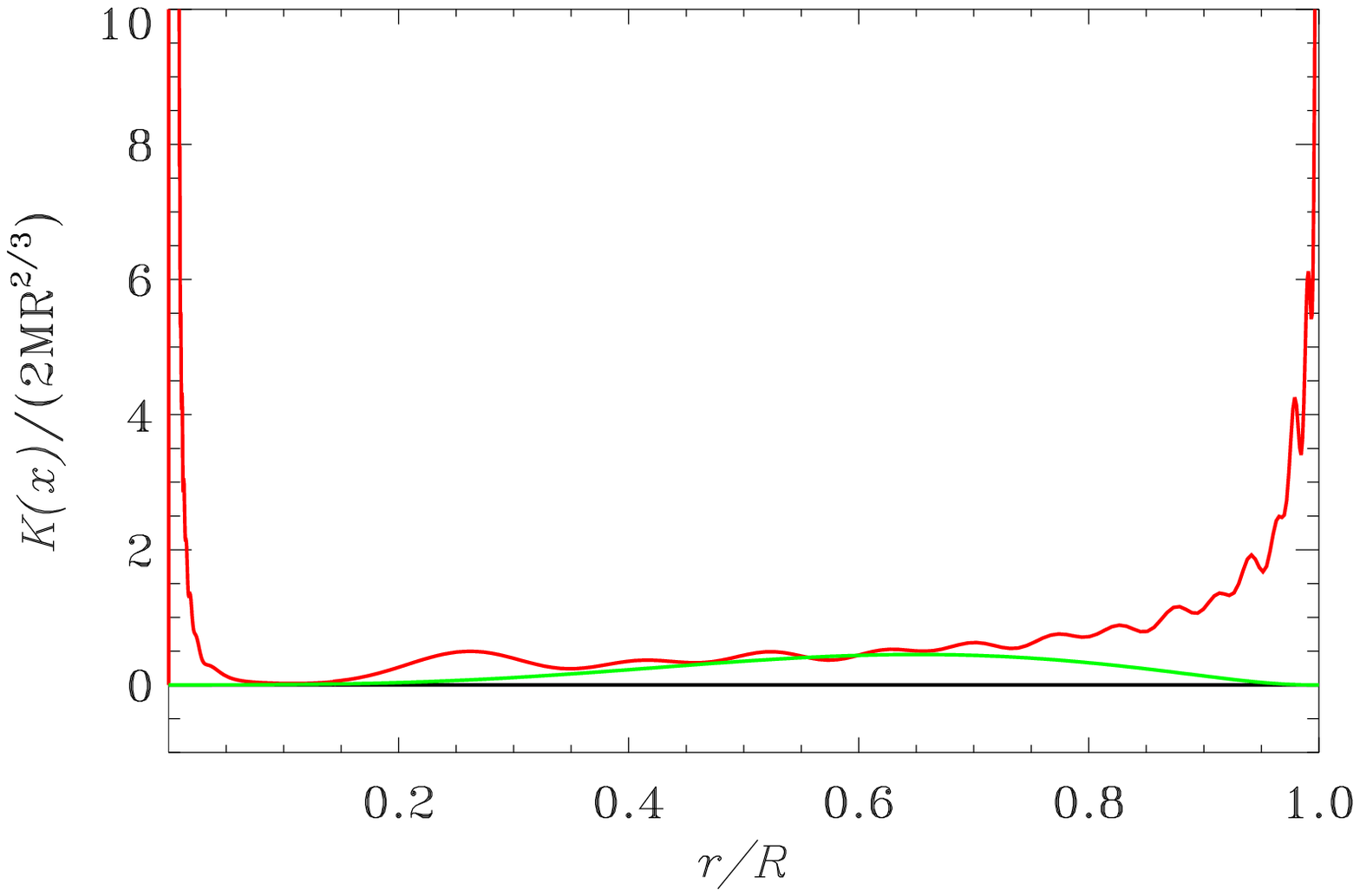}
    \includegraphics[width=9cm]{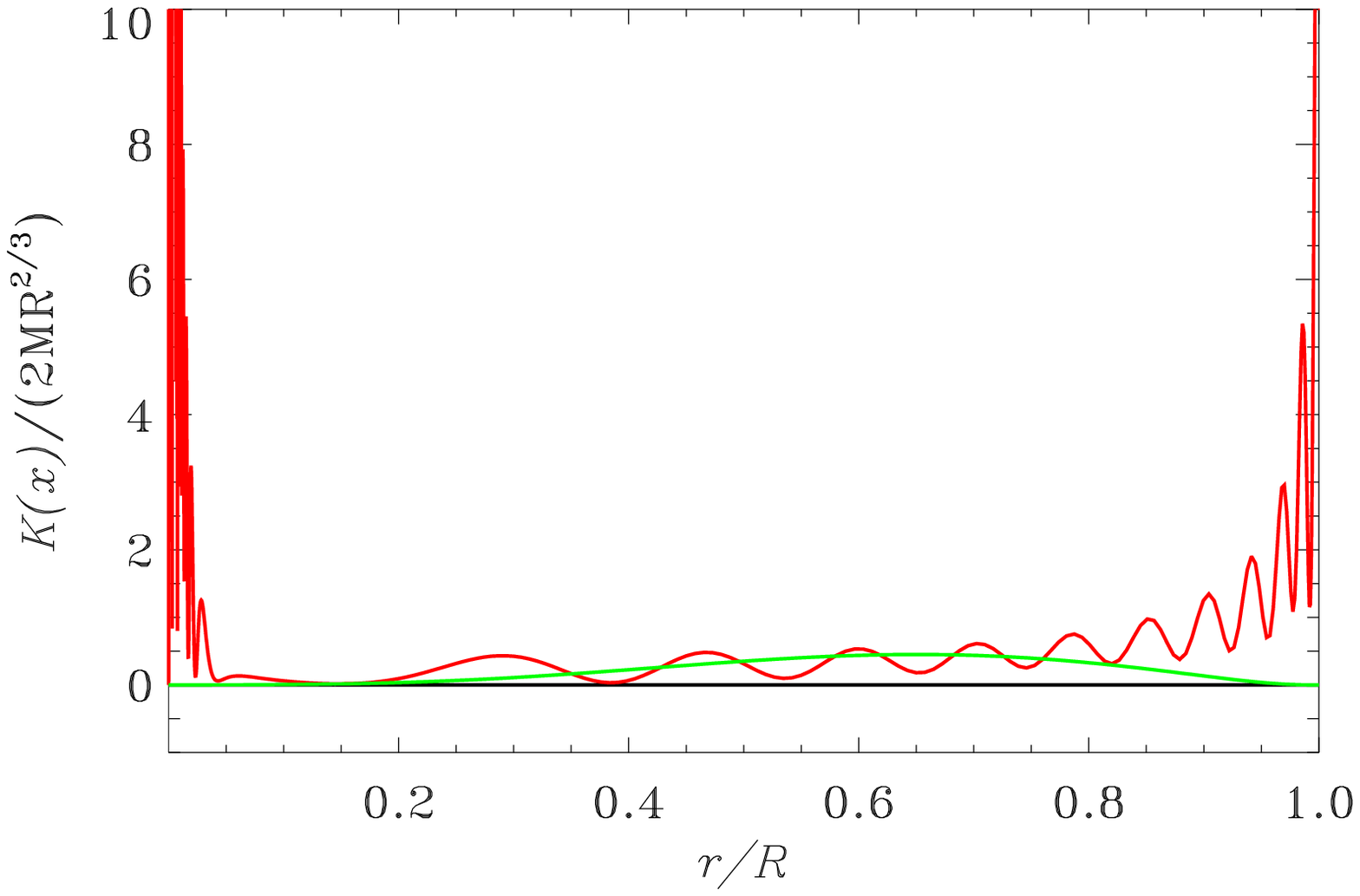}
  \caption{Averaging resolving kernel (in red) plotted in comparison with the target kernel (in green) for the inversion of a
 a set of 20 artificial splittings with $l=1$ calculated for the best-fit model of the red giant 
 KIC~4448777. Top panel: Inversion of artificial data with errors equal to 15\,nHz and a trade-off parameter $\mu=140$.
 Bottom panel: inversion of artificial data with real errors and $\mu=3800$. }
    \label{kernels2}
\end{figure}

In order 
to evaluate the weight of the errors affecting the observed splittings,
in a second 'hare-and-hounds' exercise we inverted the same set
of artificial splittings, but adopting real errors as those affecting the observed data set of 
KIC~4448777 \citep{DiMauro2018}.
By varying the trade off parameter between $100\leq \mu\leq 8000$, we obtained for $\mu=3800$ the value of $\overline{J_{(3)}}=(1.47\pm 0.02) \cdot 10^{48} \mathrm{g\, cm^2\, s^{-1}}$, which for the inversion of this set represents the closest value and hence the best agreement with ${J}_{(1)}$, characterised by $\chi^2=31$. The regularisation procedure can be visualised in Fig. \ref{chi-mu} (right panel), which shows the obtained inversion solutions $\overline{J_{(3)}}$ and the relative $\chi^2$ as functions of different values of the trade-off parameter $\mu$. 

Figure \ref{kernels2} compares averaging and target kernels for these tests.
In practice, as shown in \cite{Pijpers1994}, it is impossible to reproduce perfectly the target kernel as a consequence of the fact that  the problem is ill-posed: the set of data includes a finite number of observational constraints, the solution is not unique and in addition it is highly sensitive to noise. Nevertheless, as already explained in Sec. \ref{sec3.1},  we should not be bothered by the difference below $r=0.2\,R$ and above $R=0.9\, R$, because these regions do not substantially contribute to the measurement of the angular momentum in the star (see Sect. 2). It is interesting to point out that $\chi^2$ get closer to 1,
if we ignore the mismatch for $r<0.05 \,R$.
As it can be seen by comparing  the two panels of Fig. \ref{chi-mu} with the two panels of Fig. \ref{kernels2}, the better  precision of the data produces a better  precision in the inversion result, but real and smaller errors in the data reflect in a worse agreement between averaging and target kernels, i.e., higher $\chi^2$.

The overall good agreement found in these exercises shows that in the case of red giants, in which the dramatic structural changes due to evolution have determined the onset of mixed modes, the total angular momentum can be estimated fairly well by using sets of rotational splittings of only dipolar modes.
 The reason for that is the peculiar behaviour of mixed modes. Those of them characterised by low inertia propagate mainly in low-density layers (the acoustic cavity) of the star and behave mostly as p modes, while those characterised by a quite high inertia propagate mainly in high-density regions (the gravity-wave cavity) and behave as g modes.   

However the radial distance between the acoustic and gravity cavities, well separated in main sequence solar-like oscillators, becomes progressively smaller as the star evolves, allowing - at some stage - g modes to couple with p modes.
In red giants several g modes can couple with a single p mode of similar frequency and same harmonic degree, leading to a plethora of mixed modes per acoustic radial order with gravity characteristics in the deep interior, acoustic behaviour in the outer layers and amplitudes enhanced enough to be detectable at the surface. 
Coupling is most effective for dipole modes, while it is weaker for quadrupole and higher harmonic degree modes because of the larger evanescent zone between the two cavities. In addition, as a consequence of the smaller amplitudes at the stellar surface,  modes with $l>1$ are harder to detect.

\begin{figure*}
    \centering
    \includegraphics[width=16cm]{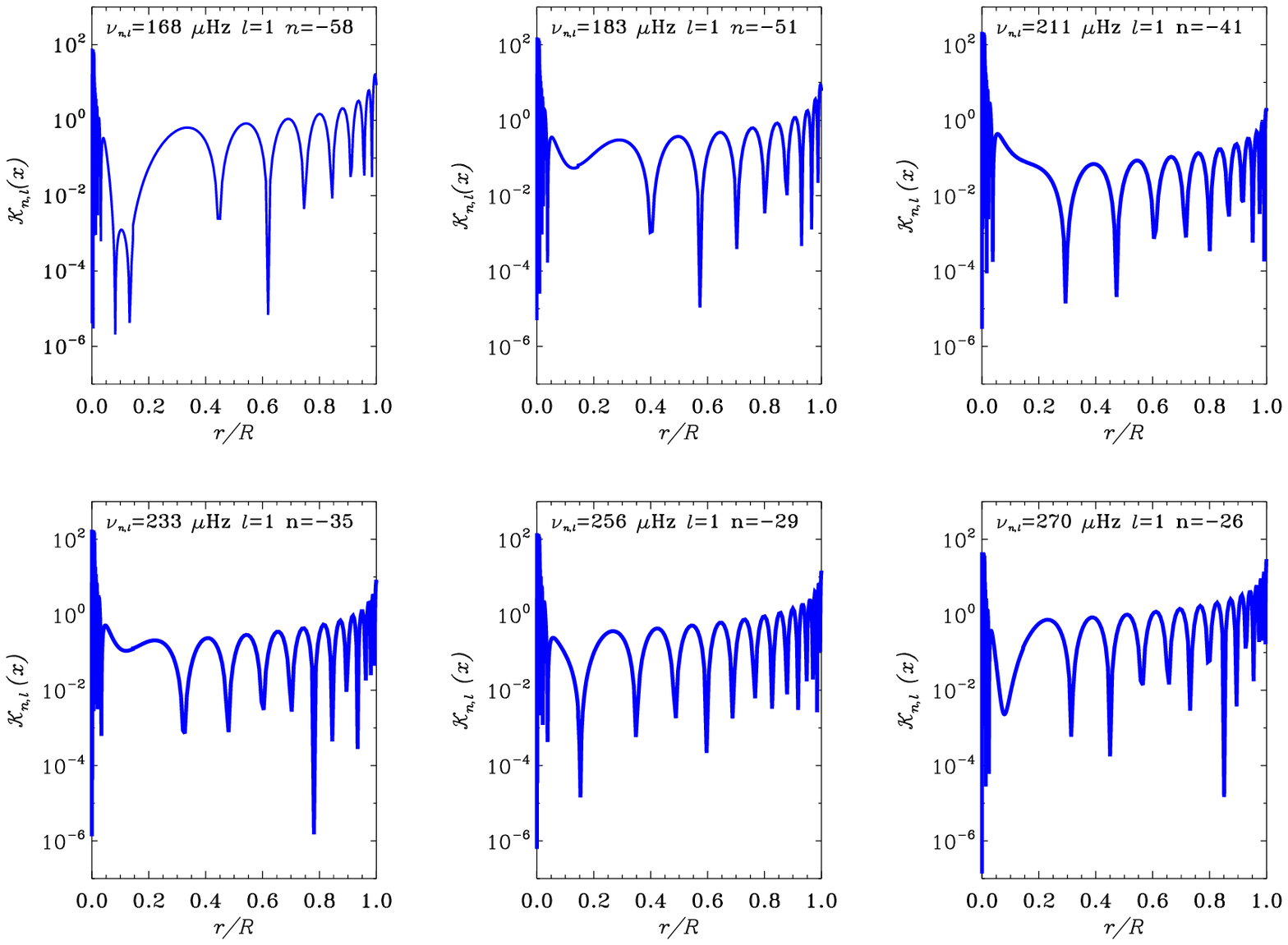}
    \vspace{-6cm}
 \caption{Individual kernels as calculated for the best-fit model of KIC~4448777 for 6 dipolar modes detected in the star. In each panel the corresponding frequency $\nu$, the harmonic degree $l$, and the radial order $n$ are indicated.}
    \label{kerl1}
\end{figure*}

A close look at the behaviour of the individual kernels of some dipolar modes, built for our best-fit model of KIC~4448777 and plotted in Fig. \ref{kerl1},  further clarify this point.
 Dipolar modes characterised by different frequencies and energies sound in different way all the internal stellar structure, from the core to the surface. 
As a consequence, in order to adequately probe the interior of a red giant it is sufficient to handle a set of modes with different gravity-acoustic character and different energy \cite[see, e.g.,][]{DiMauro2011}. 

\subsection{Improving the angular momentum determination}
\begin{figure*}
    \centering
\includegraphics[width=16cm]{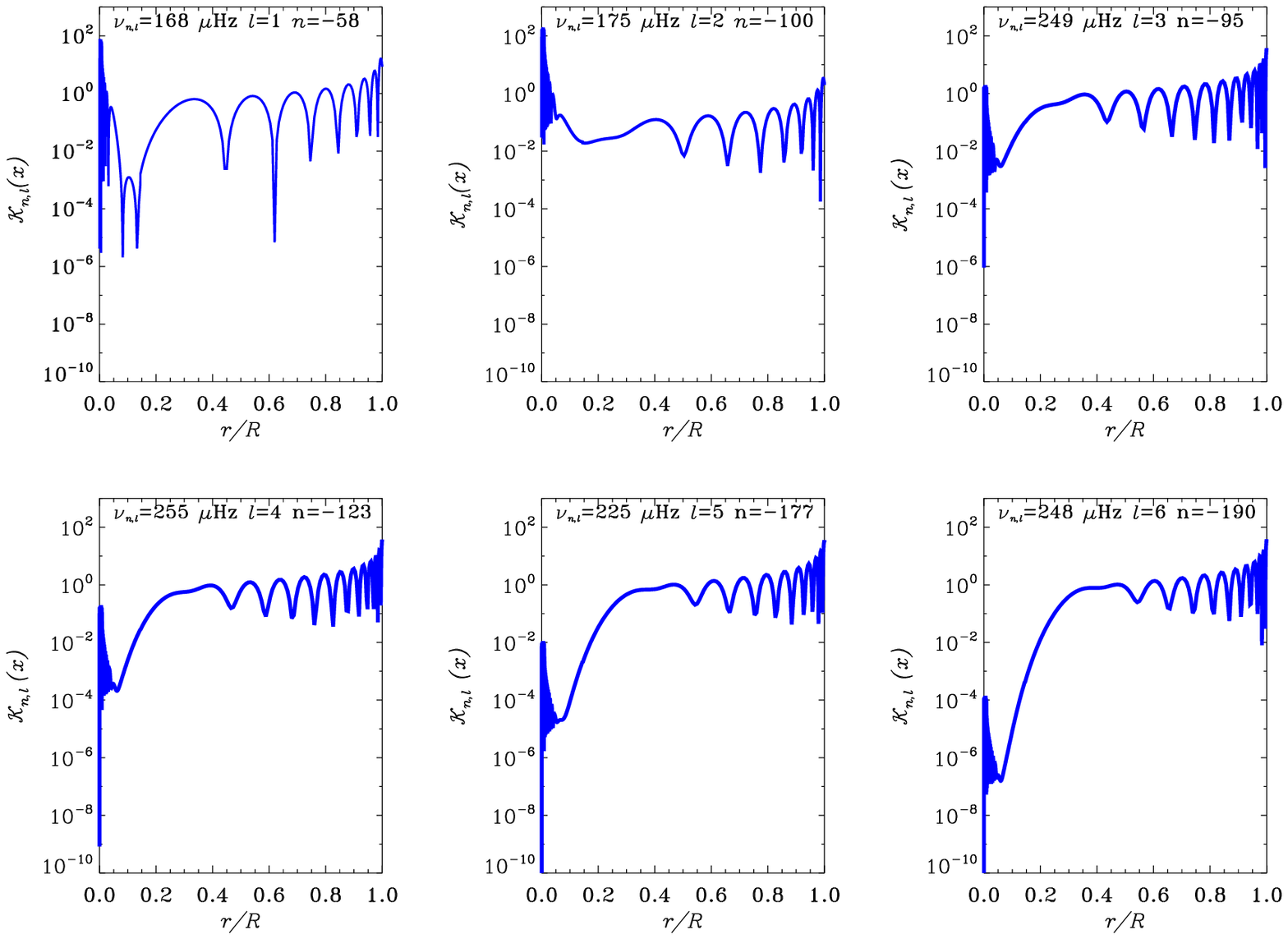}
   \vspace{-6cm}
 \caption{Individual kernels as calculated for the best-fit model of KIC~4448777 for 6 modes with increasing value of the harmonic degree. In each panel the corresponding theoretical frequency $\nu$, the harmonic degree $l$ and the radial order $n$  are indicated.}
    \label{kerl}
\end{figure*}

The results obtained  raise the natural question of whether the accuracy and the precision in the computation of the angular moment should be further improved 
by increasing the number of rotational splittings in the inverted data-set.
 To answer this question
we computed artificial splittings with harmonic degree $l\geq 2$ and frequencies ranging in  the observed interval of significant power for KIC~4448777, $160\,\mathrm{\mu Hz}\leq \nu\leq 280\, \mathrm{\mu Hz}$. An error of 15 nHz has been assumed for all the modes.
The plots reported in Fig. \ref{kerl}, obtained for the individual kernels of modes with increasing  $l=1-6$, show that the acoustic cavity moves progressively towards the surface as the value of the harmonic degree increases, causing modes with higher degree to probe more efficiently layers closer and closer to the surface. 
 At the same time for $l > 2$, the mixed modes in the interval of detected frequencies appear to show a predominant gravity nature with kernels characterised by large amplitude in the core as l increases, becoming less adequate to be used as seismic probes of the acoustic cavity.

\begin{figure}
    \centering
    \includegraphics[width=9cm]{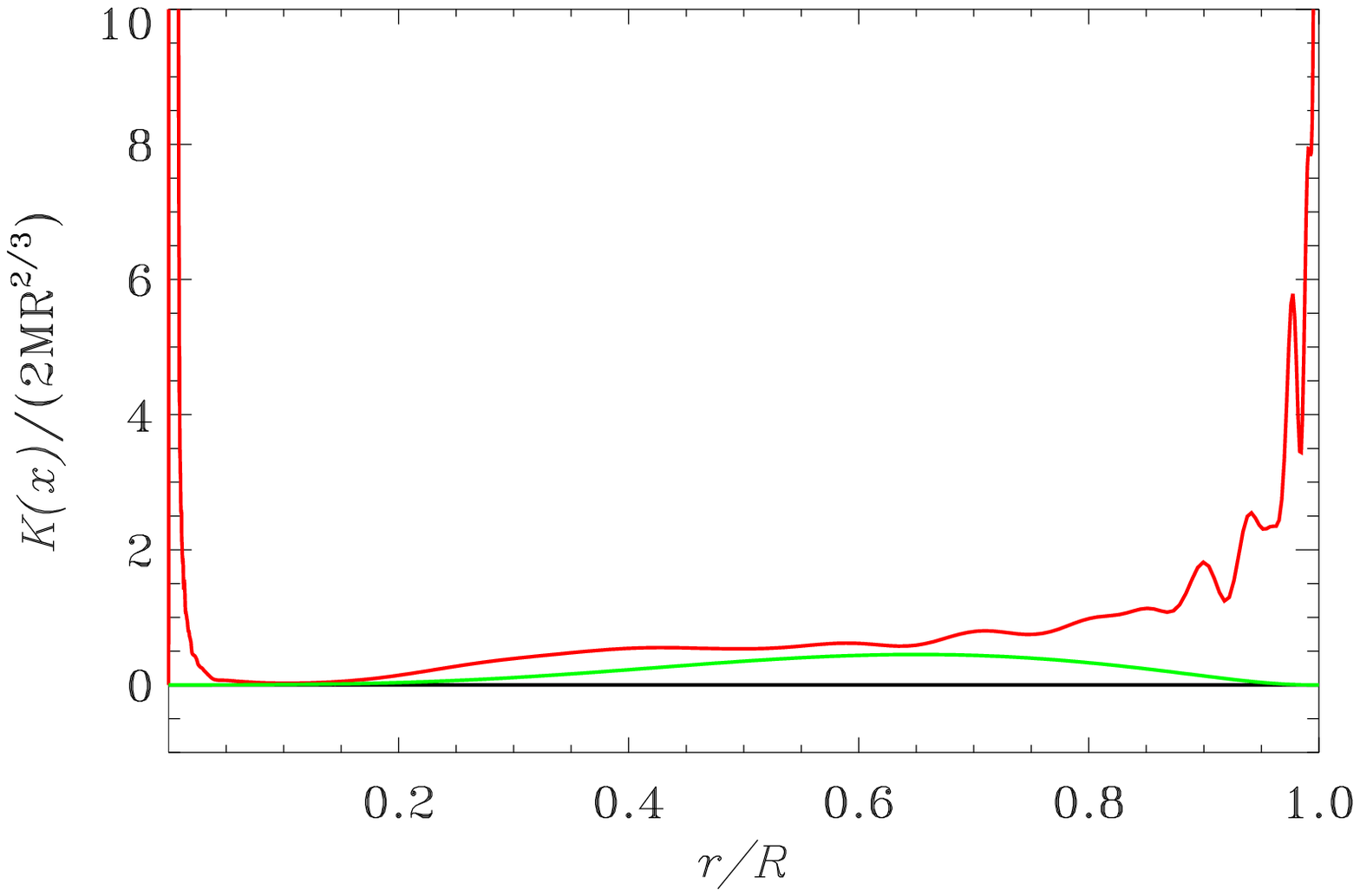}
  \caption{Averaging resolving kernel (in red) and  target kernel (in green) for the inversion of a
 set of 448 artificial splittings calculated on the best-fit model of KIC~4448777 with $l=1-6$, errors $\epsilon_i=$15 nHz and  a trade-off parameter $\mu=435.5$. }
    \label{kernels3}
\end{figure}

The comparison of the results obtained by inverting several sets of data including  mixed modes with harmonic degree $l\geq 2$ provide us the answer to the original question: in red giants the angular momentum can be estimated with an agreement between averaging and target kernels that can be only subtly improved  by considering an increasing number of splittings in the inverted data set, even using large set of data. In parallel, the precision of the result gets worse.
 For example, the result of the hare and hounds exercise for a set of data which includes 448 modes with harmonic degree $1\leq l\leq 6$
produced  $\overline{J_{(4)}}=(1.55 \pm 0.19)\cdot 10^{48} \mathrm{g\, cm^2\, s^{-1}}$ for $\mu=435.5$ and a $\chi^2=6$. This result, compared to the one obtained by inverting only 20 rotational splittings, shows clearly that no major improvement was gained by using such a large set of mixed modes,  the uncertainty gets even worse compensated by a slightly better accuracy represented by a lower $\chi^2$.
 Fig. \ref{kernels3} shows the comparison between the averaging and target kernels for the inversion of the large set of 448 artificial splittings.

It may appear counter-intuitive that adding more data, and therefore incorporating more kernels in the inversion, does not lead to improvements.
The reason can be found in the information contained in the available kernels and can be understood by considering the vector and matrix notation of the inversion problem given in Eq.~(\ref{vect}).
The eigenfunctions of the oscillation modes, from which the kernels are constructed through Eq.~(\ref{Fker}),  of course form an orthogonal complete base set.
One might infer from this that each new measured rotational splitting included in the set should add new independent information. 
However,  the eigenfunctions $W_{ij}$  of the
$(N+1)\times(N+1)$
matrix
$\mathrm{W}$ appear to  be not orthogonal, in fact some of them result to be associated to the same eigenvalue, leading to an effective reduction of the independent information which one can derive from the data set. 
Hence, one way of quantifying the information content in the available data set is to consider the eigenvalues spectrum $v_i$ of the cross correlation matrix of the rotational kernels of the set, obtained after the inversion 
(Eq.~\ref{vect}).

The eigenvalues indicate how many independent solutions the system of linear equations has and hence give information about how ill-conditioned the problem is. The  eigenvalues are scalars and are usually sorted in descending order, providing information on the reduction of the components of the new subspace for the matrix $\mathrm{W}$:
\begin{equation}
    |v_1|> |v_2|\geq |v_3|\geq |v_4|\ldots\geq |v_N|.
\end{equation}

\begin{figure*}
  \centering
  \hspace{-0.5cm}
    \includegraphics[height=8cm]{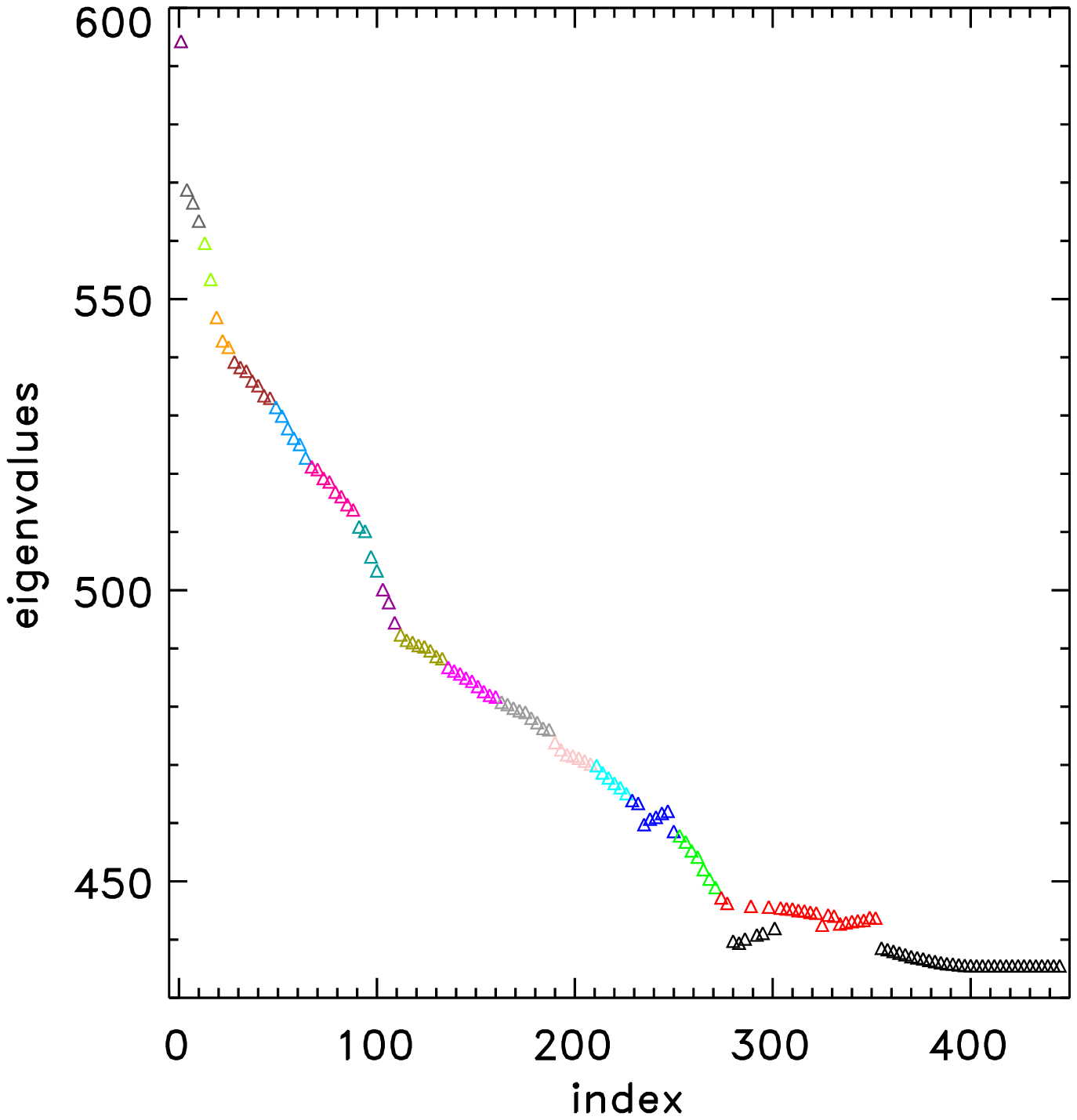}
    \hspace{-0.5cm}
    \includegraphics[height=8cm]{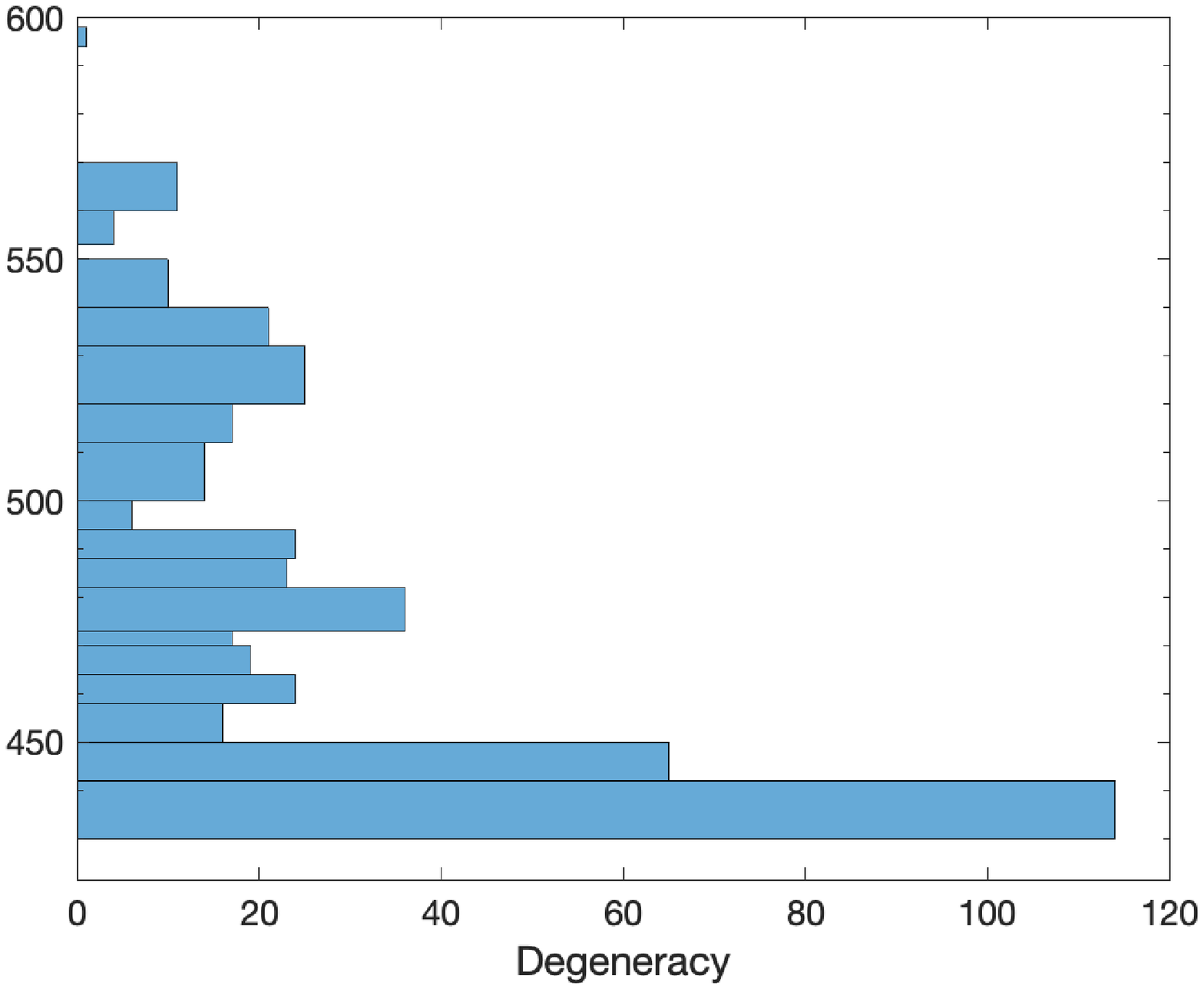}
  \caption{Left panel: Eigenvalues of the inversion matrix (Eq. \ref{vect}), plotted as function of the increasing position index, as obtained for a set of 448 artificial splittings with $l=1-6$. Different colours indicate similar solutions obtained for the system of linear equations inverted. On the right panel, the histogram shows the degeneracy of the eigenvalues, i.e. the number of times similar number appears in the vector of eigenvalues.}
    \label{histogram}
\end{figure*}

The left panel of Fig. \ref{histogram}, obtained for the adopted set of 448 artificial data, shows the resulting eigenvalues $ v_i$
plotted in order of increasing index, simply as a function of the position in the matrix, from 1 to 449.
The sequence is characterised by several plateaus,  levels of small decline or no growth,
indicating data originating same eigenvalues.
If two (or more) eigenvalues are identical, the system of linear equations is degenerate so that there are two (or more) different eigenvectors, or eigensolutions since each eigenvector corresponds to a function, which can be combined into any linear combination and also satisfy the eigensolution properties. If two eigenvalues are not perfectly identical but very close in value, as is the case here, then the corresponding eigensolution functions are normally also very similar in oscillatory character, i.e., with a very similar wavelength, but slightly different places for their maxima and minima.
The right panel of Fig. \ref{histogram} reports the degeneracy of the eigenvalues for the adopted set, showing that out of 448 modes only about 18 are contributing to add information to the inversion, while the others are adding only noise to the inversion result. This is not surprising and it was already pointed out for the case of the Sun by \citet{Pijpers1994}, who applying
the SOLA inversion technique to estimate the solar internal rotation found  that out of the 834 observed modes included in the data set only 66 of them  were really useful for the inversion.

It is well known that the degeneracy of the eigenvalues is always related to some
spatial symmetry of the system. In the present case,
this is due to the nature of the mixed modes and the way in which these modes probe the interior of red-giant stars, as discussed above. According to what we have learnt from the Sun, we would expect a different conclusion if the inverted data set could also include pure p or g modes, which unfortunately are not excited in red-giant stars.
The implication is that each extra measured rotational splitting does not add completely independent information to the set. 

Since any real measured or even  artificial  data-set  consists of a finite  number of frequency splittings, the dimension of the functional space spanned by the corresponding eigenfunctions is also limited.  Due to the degeneracy,
the dimension of the function space spanned by the kernels is even smaller, and barely increases as the number of available splittings increases. Especially when all the available modes are mixed modes, with very short spatial wavelength in the core, adding further modes in the inversion algorithm produces an averaging kernel which crosses the target kernel very often, so that at those locations the inversion appears successful, but at the same time the averaging kernel has much larger deviations from its target at other values of $r/R$. 
For any given data-set, this behaviour can only be suppressed by penalising large values for the linear coefficients, i.e. setting the regularisation parameter $\mu$ to very high values. In other words, the take-home lesson here is that diversity of the modes represented by the data is more important than merely having a large quantity of modes.

\subsection{Angular momentum by applying the SOLA method}

 Since the properties of the inversion  depend, for a given stellar model, on the errors in the rotational splittings and the modes included in the data-set, the trade-off parameter $\mu=3800$ found with the hare and hounds test in Sec. 3.2, can be adopted for the inversion of the observed set of data.
The result obtained by applying the SOLA method to the 
red-giant star  KIC~4448777 indicates a value of the total angular momentum  $ \overline{J_\star}\pm \Delta J_{ \mathrm {stat}}= (3.90\pm  0.02) \cdot 10^{48}\,\mathrm{ g\,cm^{2}\,s^{-1}}$, value which is in good agreement with $J_{\mathrm{tot}}$ obtained in Sect. 3.1, but much better constrained.
The percentage error
$ \Delta J_{\mathrm {stat}}/J_\star=0.5\%$  corresponds only to  the statistical uncertainty coming from the errors propagation in the inversion procedure and is related only to the observed frequency splitting uncertainties.

In order to get a more realistic estimate of the angular momentum, we need to consider the contribution to the global error coming from the accuracy with which the selected stellar model best-fits the observations of the star.
 It was demonstrated by \cite{Deheuvelsetal2014} that no significant difference is found in the inversion results for the internal rotation by using different stellar models, as long as these models are consistent with each other to
first order, which means that they are able to match within the errors both seismic and non-seismic parameters. \citet{DiMauro2016} showed that  two selected best-fit models, chosen on the basis of the $\chi^2$ criterion, produce similar rotational inversion results. The good agreement of the  results  has been confirmed also by applying different independent methods (i.e., the method proposed by \citealp{goupil2013} and  a least-squares fit to the observed rotational splittings).

\cite{Reese2015} explained that
some discrepancy due to the use of different models rise from the fact
 that different models might have oscillation modes
with different inertia because of possible different extents of the acoustic and gravity cavities . In the case of red giants, the regions above the core are mostly sounded by modes with mixed g–p character, whose identification is more crucial than modes with dominant p or g behaviour, respectively, better
trapped in the convective region and in the inner core  \cite[see, e.g.,][]{Mosseretal2011, Mosseretal2012}.
Fortunately, mixed modes of 
different inertia have very
different damping times and profiles in the observed oscillation spectrum \cite[see, e.g.,][]{Mosser2018}. Thus, the analysis of the observations \citep[see, e.g.,][]{Corsaro2014} can provide essential information on the gravity or acoustic nature of each detected oscillation mode and the best-fit stellar model is chosen by using the nature of each mode, not only the oscillation frequencies, as additional constrains.

Furthermore, \cite{Schunkeretal2016} explored the sensitivity of the inversion procedure to different stellar models and they concluded that for Sun-like stars with moderate radial differential rotation gradients the inversions are insensitive to uncertainties in the stellar models.

In order to quantify the dependence of the result on the assumed stellar structure model, we can simply consider that the angular momentum depends on the moment of inertia - given roughly by $I \propto M R^2$ -  i.e. on the distribution of mass inside the stellar models and also on the total mass and radius of the star. 
Considering that the  accuracy level of asteroseismic estimates by using only the average seismic parameters for KIC~4448777 as calculated in \cite{DiMauro2016} and shown in Table \ref{tab_fitted} is 
$\Delta M/M=8\%$ for the  stellar mass   and  
$\Delta R/R=3\%$ for the stellar radius, we obtain that in order to take into account  the uncertainty in model computations, we need to consider an additional relative systematic error of about
$\Delta J_{\mathrm{syst}}/J_{\star}= 14\%$,
 corresponding to an uncertainty  $ \Delta J_{ \mathrm{syst}}= 0.55 \cdot \,  10^{48} \mathrm{g\, cm^2\, s^{-1}}$.

We can conclude that  the total angular momentum of KIC~4448777
can be evaluated  by the asteroseismic technique and it is: 
\begin{equation}
\overline{J_\star}\pm\Delta J_\star=\overline{J_\star}\pm(\Delta J_{ \mathrm {stat}}+ \Delta J _{\mathrm{syst}})
=(3.90 \pm0.57) \cdot \, 10^{48} \mathrm{g\, cm^2\, s^{-1}}\, ,
\end{equation}
with a total  relative uncertainty $\Delta J/J_\star=14.5\%$. This result is an enormous  improvement with respect to the  method of integrating Eq.~(\ref{Eq.3}), based on the known internal rotational profile, despite the relatively small set of  rotational splittings,  covering a limited range in $l$, available for the inversion.

Finally, we tested the above estimate of the total uncertainty by adopting 
 a different evolutionary model well fitting, within the errors, both seismic and non-seismic
 observations of KIC~4448777, but with a lower goodness of the chi-square test 
 as described in \cite{DiMauro2018}. The second-best model is characterised by the following parameters: $M=1.02\,{\mathrm M}_{\odot}$, $T_{\mathrm{eff}}=4800$\,K, $R=3.94{\mathrm R}_{\odot}$, $\log g= 3.26$\,dex. Once selected a different best trade-off parameter for this model by performing a new hare and hounds test,
 we obtained for $\mu=2160$ a value 
 \begin{equation}
  \overline{J_\star}\pm\Delta J_\star={\overline{ J_{\star_2}}}\pm (\Delta J_{\mathrm {stat}}+ \Delta J _{\mathrm{syst}})=(2.94 \pm 0.43) \cdot \, 10^{48} \mathrm{g\, cm^2\, s^{-1}}
 \end{equation}
 and $\chi^2=29$.
 The reported uncertainty includes the statistical error $\Delta J_{\mathrm {stat}}=0.02$ coming from the inversion procedure, merely related to the observed frequency splitting uncertainties, and the systematic error $\Delta J _{\mathrm{syst}}=0.41$ arising from uncertainty in the stellar mass and radius of the adopted model.
 
 We can conclude that the results $\overline{J_\star}$ and $\overline{J_{\star_2}}$ obtained by using the two different best-fitting models are in agreement within the errors and hence we can confirm  that the procedure is model independent within the limits described above.

\section{Discussion and conclusion}
This paper demonstrated that the total angular momentum of red-giant stars can be well and properly determined through the SOLA asteroseismic inversion technique.  More interestingly, this holds even if a relatively limited  set of data including rotational splittings of only dipolar modes, is available.

The usual route is to first determine an average internal angular velocity or 'as best as possible' a spatially resolved rotation rate  and then reintegrate that, after multiplying by the moment of inertia calculated for the adopted 'best' evolutionary model of the star.
This route has unfortunate error propagation properties, with the result that the uncertainty in the angular momentum, due to the measurement errors of the data, becomes quite large. 

Here we have analysed the inference of the total angular momentum of KIC 4448777, a red-giant star of stellar mass $M=(1.12\pm0.09)\,\mathrm{M}_{\odot} $, in which observations performed by the space mission {\it Kepler} have allowed to detect 20 rotational splittings of oscillations frequencies of harmonic degree $l=1$.
By using an appropriate target kernel within the SOLA formalism, described here, the angular momentum of this star has been determined directly from the inversion of  splittings, taking the full 2-dimensional character of the internal rotational profile into account. We find that the SOLA inversion procedure is characterised by good error propagation properties.
This allows us to obtain for our star a total angular momentum 
$
\overline{J_\star}\pm(\Delta J_{ \mathrm {stat}}+ \Delta J _{\mathrm{syst}})=[3.90 \pm(0.02+0.55)] \cdot \, 10^{48} \mathrm{g\, cm^2\, s^{-1}}
$
 where the total uncertainty
includes the statistical error rising from the data and the systematic error due to the accuracy in mass and radius of the star.
For KIC~4448777, characterised by very  precise observational data, it is just this last source of error which dominates the total uncertainty. This will generally be an issue for any star other than the Sun.

Moreover, it should not be a surprise that, even if asteroseismic inversions  \citep{DiMauro2016,DiMauro2018} have produced radial profile of the angular velocity of KIC~4448777 characterised by low resolution in the convective envelope (see low panel of Fig. \ref{angmom}), the global angular momentum results are not affected by such large errors. 
In fact, the SOLA inversion method in the present work has been applied not to search for localised  solutions along the stellar radius, but to determine a global parameter.

In addition, it has been demonstrated that in order to determine
the total angular momentum of red-giant stars with  good precision and accuracy, it is sufficient to use data sets which include rotational splittings of only dipolar modes, as those already obtained for several targets by the photometric space missions  and a model which best-fits the seismic and non-seismic observations of the star. 
The contribution to the analysis of  splittings of high-harmonic degree modes seems to not improve the precision of the inversion,  since we have demonstrated that the diversity of oscillation trapping character, which translates in modes with different inertia, 
is more important than the employment of just a large number of modes. 
It is doubtless that a large data set would contribute to better constrain the best-fit model necessary for the inversion procedure and then improve the accuracy of the result.

 In the present manuscript we did not explore how changes to the input
  stellar models might affect the inversion results. Nevertheless, the precision of the actual asteroseismic data-sets demands that the sources of systematic errors has to be addressed. Hence, it will be worthwhile in the future
 to carry out a systematic study devoted to investigate how the choice of the best-fit model might impact the inversion of angular momentum, as it was done previously for the determination of the angular velocity.
This can be done by employing different evolutionary codes with different input physics
regarding diffusion, core overshoot and rotation, even though the
largest source of error  in all the asteroseismic studies unfortunately still arises from modelling the near-surface layers \citep[][and references therein]{jorgensen2021}.
This procedure will provide a smaller systematic error than we currently estimate.

For the near future the authors plan
 to extend this technique to the large
sample of solar-like
 stars for which rotational splittings of dipolar oscillation modes have been successfully
 detected  by {\it Kepler}/K2.  In particular it will be interesting to study
the evolution of the stellar
total angular momentum
from main sequence to later stages, by selecting observed stars with  similar stellar masses, but in different evolutionary phases.
 This will allow to understand the dependence of the angular momentum on the 
 age of the star and possibly to estimate the amount of angular momentum transferred from the core to the envelope during the evolution in order to balance the rotational velocity.

Moreover, we will try to understand if
main sequence solar-like stars with detectable rotational splittings  \citep{Benomaretal2015}, and to some extent also subgiants, might represent better subjects in this context. In these stars the target kernel and the achievable averaging kernel might show a better match, due to the use of observed data-sets, which include mainly acoustic modes with low harmonic-degree trapped in the regions relevant for the evaluation of their total angular momentum. 

Furthermore, we plan to apply this technique even to intermediate- or high-mass stars, 
in which tens of rotational splittings of g and p modes have been already detected \citep{Aerts2019}.

We believe that our tool based on a data-driven approach,  in which very accurate space mission observations act as a test bench for theoretical simulations and models, will help to understand fundamental processes governing
the variation of the internal rotation of stars with time, such as the
core-envelope coupling or the angular momentum loss from the
surface.
The estimate of the total angular momentum of a star, incorporated as an essential ingredient
or as a boundary condition into 
rotating stellar evolutionary codes, will produce the needed advancement of the theory of angular momentum transport in stellar interiors.

The limit of application of this technique might be represented by the difficulty to measure rotational splittings of oscillation modes with too low amplitude to be detected with sufficient accuracy or hidden
in a forest of multiple modes of oscillations in different kind of stars.
However, the space missions NASA/TESS \citep{Rickeretal2014}, launched in April 2018, and the 
ESA/PLATO \citep{Raueretal2016}, to be launched in the 2026, have the potential to enable us to detect rotational splittings and hence to infer the total angular momentum in a large sample of stars, including high-mass and metal-poor stars in binaries and clusters.

 \begin{acknowledgements}
 \\  The authors thank very much  the anonymous referee  for
 his/her useful suggestion and comments, which gave the opportunity to
greatly improve the manuscript.
 \end{acknowledgements}

\bibliographystyle{aa} 
\bibliography{angular} 
\end{document}